\documentclass[conference]{IEEEtran}

\PassOptionsToPackage{table}{xcolor}
\usepackage{cite}
\usepackage{amsmath,amssymb,amsfonts}
\usepackage{algorithmic}
\usepackage{graphicx}
\usepackage{textcomp}
\def\BibTeX{{\rm B\kern-.05em{\sc i\kern-.025em b}\kern-.08em
    T\kern-.1667em\lower.7ex\hbox{E}\kern-.125emX}}
\usepackage{todonotes}
\usepackage{url}
\usepackage{xspace}
\usepackage{soul}
\usepackage{threeparttable}
\usepackage{enumitem}
\usepackage{booktabs}
\usepackage{multirow}
\usepackage[table]{xcolor}
\usepackage{colortbl}

\newcommand{\tool}{\textsc{Mutabot}\xspace}

\begin{document}

\title{Towards Multi-Platform Mutation Testing of Task-based Chatbots\\
}

\author{
\IEEEauthorblockN{Diego Clerissi}
\IEEEauthorblockA{\textit{University of Milano-Bicocca} \\
Milan, Italy \\
diego.clerissi@unimib.it}
\and 
\IEEEauthorblockN{Elena Masserini}
\IEEEauthorblockA{\textit{University of Milano-Bicocca} \\
Milan, Italy \\
elena.masserini@unimib.it}
\and
\IEEEauthorblockN{Daniela Micucci}
\IEEEauthorblockA{\textit{University of Milano-Bicocca} \\
Milan, Italy \\
daniela.micucci@unimib.it}
\and
\IEEEauthorblockN{Leonardo Mariani}
\IEEEauthorblockA{\textit{University of Milano-Bicocca} \\
Milan, Italy \\
leonardo.mariani@unimib.it}
}

\maketitle

\begin{abstract}
Chatbots, also known as conversational agents, have become ubiquitous, offering services for a multitude of domains. 
Unlike general-purpose chatbots, task-based chatbots are software designed to prioritize the completion of tasks of the domain they handle (e.g., flight booking). 
Given the growing popularity of chatbots, testing techniques that can generate full conversations as test cases have emerged. Still, thoroughly testing all the possible conversational scenarios implemented by a task-based chatbot is challenging, resulting in incorrect behaviors that may remain unnoticed. 
To address this challenge, we proposed \tool, a mutation testing approach for injecting faults in conversations and producing faulty chatbots that emulate defects that may affect the conversational aspects. In this paper, we present our extension of \tool to multiple platforms (Dialogflow and Rasa), and present experiments that show how mutation testing can be used to reveal weaknesses in test suites generated by the Botium state-of-the-art test generator.
\end{abstract}


\section{Introduction} \label{sec:introduction}
As technology advances and services are increasingly accessible, chatbots have become ubiquitous in everyday activities, being able to support users across a wide range of domains~\cite{adamopoulou2020chatbots}.
Unlike general-purpose chatbots (e.g., ChatGPT~\cite{chatgpt}), those developed to perform specific tasks (e.g., booking a hotel room, creating a Google Calendar event, or providing weather updates) are commonly referred to as \emph{task-based chatbots}~\cite{grudin2019chatbots,adamopoulou2020chatbots}.
Task-based chatbots can be implemented using a variety of platforms, including Google Dialogflow\cite{dialogflow}, Rasa~\cite{rasa}, and Amazon Lex~\cite{amazon-lex}.

Despite their growing adoption, ensuring the reliability of task-based chatbots remains a largely open challenge~\cite{lambiase2024motivations,li2022review}, as it requires the generation of conversational scenarios that exercise relevant behaviors, as well as the definition of oracles able to accurately assess the correctness of the responses. 
Botium~\cite{botium} is a state-of-the-practice testing framework for conversational agents, supporting automated test generation and execution. In Botium, a test case corresponds to a sequence of user-bot interactions, testing a chatbot's functionalities from a conversational aspect. Botium's capabilities have been leveraged by follow-up approaches, such as Charm~\cite{bravo2020testing} and CTG~\cite{rapisarda2025test}. Still, these techniques exhibit weaknesses in terms of input space and oracle precision, resulting in limited coverage of conversations and bug detection. 

Mutation testing~\cite{jia2010analysis} has thus been recently adapted to the context of chatbots~\cite{ferdinando2024mutabot,gomez2024mutation}. In this context, artificial faults (namely \textit{mutants}) 
are introduced to take into account the peculiarities of conversational aspects and development platforms, to evaluate the effectiveness of existing chatbot testing techniques in detecting these faults (i.e., \textit{killing the mutants}, according to the standard terminology~\cite{jia2010analysis}).
%
For example, Dialogflow chatbots include both JSON files, which specify user utterances (i.e., user-provided inputs) and chatbot responses, and Javascript code, which triggers custom functions. Instead, Rasa implements conversations both with multiple YAML files, which define the context domain, training data, and flow rules, and with custom actions defined as Python functions. 

We recently presented \tool~\cite{ferdinando2024mutabot}, a mutation testing tool designed to support multi-platform mutations for task-based chatbots, originally developed for Dialogflow. In this paper, we describe how we adapted \tool to the Rasa platform, reporting some preliminary findings on the effectiveness of tests generated by Botium against mutants generated with \tool. Results suggest that more research is needed to generate fault-revealing conversations.

The paper is structured as follows. Section~\ref{sec:tool} introduces the tool and the design advancements. Section~\ref{sec:experiment} discusses our findings following the experiments in Rasa. Then, Section~\ref{sec:related} discusses the related work. Finally, Section~\ref{sec:conclusion} provides some conclusions and outlines future work. 

\section{Mutation Testing of Chatbots with \tool} \label{sec:tool}
\begin{figure*}[t!]
 \centering
 \includegraphics[trim=0.0cm 0.0cm 0.0cm 0.0cm, clip=true, width=17.0cm]{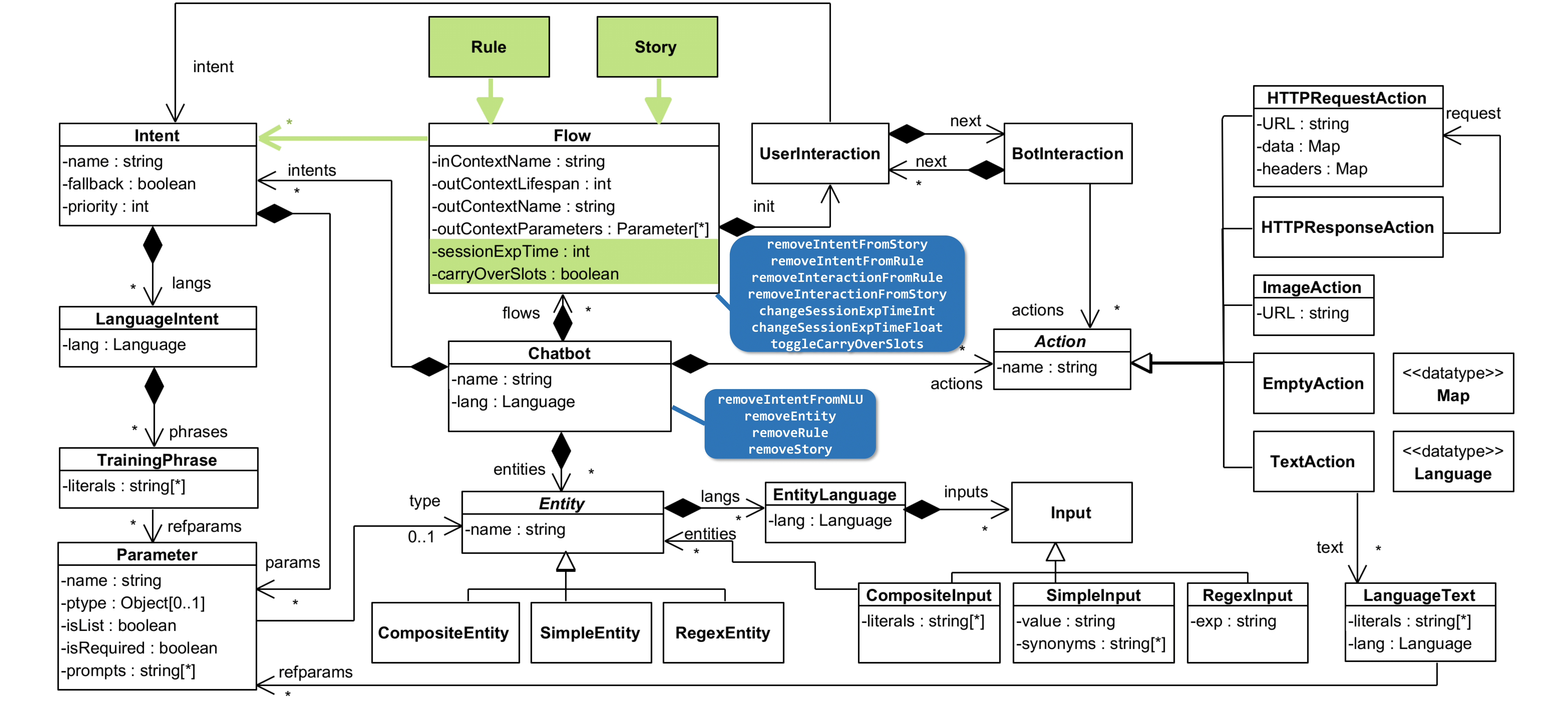}
 \vspace{-3mm}
 \caption{Chatbot structure meta-model adapted from Ca{\~n}izares \textit{et al.}~\cite{canizares2022automating}.}
 \vspace{-3mm}
 \label{fig:meta-model}
 \end{figure*}

Task‑based chatbot platforms rely on a set of key concepts that describe how conversations are structured and managed. These concepts are captured in the meta‑model shown in Figure~\ref{fig:meta-model}, adapted from the one originally proposed by Cañizares \textit{et al.}, which provides a platform‑agnostic perspective on chatbot structure~\cite{canizares2022automating}.
%
%
In particular, the meta‑model identifies the following core elements:
the \textit{intents} (i.e., the goals a user wants to achieve by interacting with the chatbot), \textit{entities} (i.e., the data types that identify the parameters used in the conversations), \textit{parameters} (i.e., also called \textit{slots} in Rasa, the variables that store the input values), \textit{actions} (i.e., the operations a chatbot performs to fulfill a user’s intents), and \textit{flows} (i.e., the user-bot conversational scenarios). 

In Rasa, flows are represented as sequences of intents and actions described through \textit{stories} (i.e., flexible conversation examples) and \textit{rules} (i.e., constrained predefined paths).

Unlike traditional mutation techniques that typically manipulate code-level constructs (e.g., conditional nodes, array indexes), \tool is designed to tackle the complexity and peculiarities of conversational agents implemented according to the aforementioned meta-model.

For example, a mutation operator can be designed to change the conversational flow of a chatbot. This operator has different implementations, depending on the target platform. In Dialogflow, flows can be manipulated by changing context objects in JSON files, while in Rasa flows can be manipulated by modifying the structure of the YAML files that describe conversational stories. The faults injected by \tool represent realistic errors that may arise from imperfectly designed chatbots, such as parameter names accidentally swapped in a response, data not correctly propagated between intents, or required entities not correctly specified in the training data.

Figure~\ref{fig:meta-model} highlights in green the additional elements we have introduced into the meta‑model 
to support Rasa, showing the names of the supported operators in the callout texts. The used meta-model includes the changes proposed by Urrico \textit{et al.}~\cite{ferdinando2024mutabot}. 
To this end, we have currently implemented the following eleven operators: (1) \textit{removeIntentFromNLU}, (2) \textit{removeEntity}, (3) \textit{removeRule}, (4) \textit{removeStory}, (5) \textit{removeIntentFromStory}, (6) \textit{removeIntentFromRule}, (7) \textit{removeInteractionFromRule}, (8) \textit{removeInteractionFromStory}, (9) \textit{changeSessionExpTimeInt}, (10) \textit{changeSessionExpTimeFloat}, and (11) \textit{toggleCarryOverSlots}. 

The first four operators affect the chatbot structure, operating on different files, by removing conversational elements, for example, the removal of an intent from the training data (operator 1) or the removal of an entity (operator 2). 
The remaining seven operators affect the flow of a conversation. For instance, operators 9 and 10 extend or shorten the lifespan of a user-bot session by a numeric value, while \textit{toggleCarryOverSlots} (operator 11) enables the reset mode for data shared among intents when a session ends. 
Whenever an operator can be applied multiple times to a chatbot (e.g., operator \textit{removeIntentFromNLU} applied to a chatbot that includes several intents), \tool iteratively applies it to all possible cases and produces distinct mutants (e.g., one mutant for each removed intent).

We speculate that some other not yet implemented mutations affecting the conversational features defined in the meta-model may not apply to all cases, or may require adaptations. For example, the intent priority, which defines the order of competing intent activations, is an explicit numeric property associated with each intent in Dialogflow, while in Rasa this property can be manipulated only indirectly by modifying the policy declaration that defines the intent confidence.  

\section{Early Empirical Evidence} \label{sec:experiment}

To conduct a preliminary investigation of mutation testing for 
Rasa chatbots, we used \tool to generate mutants from three third-party Rasa chatbots selected from the BRASATO dataset~\cite{masserini2025brasato}. To evaluate the usefulness of the generated mutants, we created test cases for each chatbot using Botium.
Table~\ref{tab:chatbots} shows the main characteristics of the subject chatbots  (i.e., their name, domain, and the number of intents/entities/actions defined) as well as the size of the test suites generated by Botium.

\renewcommand{\arraystretch}{1.2}
\setlength{\tabcolsep}{3pt}
\begin{table}[t!]
\caption{Rasa Subject Chatbots.}
\vspace{-1mm}
\label{tab:chatbots}
\centering
\footnotesize
\begin{threeparttable}
\begin{tabular}{l l c c c c}
    \hline
    \textbf{Name} & \textbf{Domain} & \textbf{\# Int.} & \textbf{\# Ent.} & \textbf{\# Act.} & \textbf{\# Tests} \\
    \hline
    Rock Paper Scissors\tnote{a} & Entertainment & 6 & 1 & 1 & 46 \\  
    PJs Chatbot\tnote{b}         & Food \& Drink & 7 & 8 & 4 & 74 \\
    Customer Service\tnote{c}    & Business      & 20 & 2 & 18 & 83 \\
    \hline
\end{tabular}
\begin{tablenotes}\footnotesize
    \item[a] \url{https://github.com/naveedeveloper/RASA-RPS-Challenger}
    \item[b] \url{https://github.com/ChristianCitterio/pjs_chatbot_rasa}
    \item[c] \url{https://github.com/farhadmohmand66/customer_care_chatBot}
\vspace{-4mm}
\end{tablenotes}
\end{threeparttable}
\end{table}

\newcolumntype{g}{>{\columncolor[gray]{0.9}}c}
\renewcommand{\arraystretch}{1.2}
\setlength{\tabcolsep}{6pt}
\begin{table*}[ht!]
\centering
\caption{Mutants killed over total by Botium-generated test suites.}
\vspace{-1mm}
\begin{threeparttable}
\begin{tabular}{l|ccccc|ccccc|ggggg}

\hline


\multirow{2}*{\textbf{Chatbot}} & \multicolumn{5}{c|}{\textbf{Chatbot}} 
& \multicolumn{5}{c|}{\textbf{Flow}} 
& \multicolumn{5}{>{\columncolor[gray]{0.9}}c}{\textbf{Total}} \\
& \textbf{B} & \textbf{K} & \textbf{E} & \textbf{G} & \textbf{\%K}
& \textbf{B} & \textbf{K} & \textbf{E} & \textbf{G} & \textbf{\%K}
& \textbf{B} & \textbf{K} & \textbf{E} & \textbf{G} & \textbf{\%K} \\
\hline
Rock Papers Scissors 
& 0  & 6 & 0 & 12 & 50\%
& 0 & 6 & 2 & 15 & 47\%
& 0 & 12 & 2 & 27 & 48\% \\
\hline
PJs Chatbot 
& 0 & 20 & 3 & 26 & 87\%
& 13 & 7 & 3 & 28 & 58\%
& 13 & 27 & 6 & 54 & 77\% \\
\hline
Customer Service
& 0 & 23 & 11 & 48 & 62\% 
& 12 & 8 & 7 & 54 & 23\%
& 12 & 31 & 18 & 102 & 43\% \\
\hline

\end{tabular}
\begin{tablenotes}\footnotesize
\item[*] \# Broken (B), \# Killed (K), \# Equivalent (E), \# Generated (G), \% Killed (\%K), where \%K = K / (G - B - E) * 100  

\vspace{-3mm}
\end{tablenotes}
\end{threeparttable}
\label{tab:results}
\end{table*}

For each chatbot, we first generated the Botium test cases representing our baseline regression test suite and ran them five times to account for any flakiness\footnote{A flaky test is a test that both passes and fails periodically without any code
changes~\cite{zheng2021research}.}. Then, we applied \tool to each chatbot, generating mutants based on the supported operators for Rasa. We manually inspected the chatbots and discarded the equivalent ones\footnote{An equivalent mutant is a mutant whose behavior is the same as the original software and thus cannot be killed by any test case~\cite{jia2010analysis}.}. Most equivalent mutants were caused by operators modifying the flow structure, whose effects may be nullified when an intent is removed from one flow but is still used in others (e.g., a greet intent reused across multiple flows). Additionally, we observed that the original implementation of ``PJs Chatbot'' includes a faulty action that is never executed, making related mutations equivalent by design.
Finally, we separately deployed each mutant on the Rasa platform and ran the Botium test suite, collecting the test outcomes. Results are reported in Table~\ref{tab:results}.

Botium was able to detect between 43\% and 77\% of the mutants. The mutants easier to detect were those involving elements removal, in particular intent removal, confirming the results obtained in our previous study on Dialogflow chatbots~\cite{ferdinando2024mutabot}.
This can be explained by the fact that Botium is designed to assert on intent detection: when the expected intent is missing and the behavior falls to another intent, the mutant can be easily killed.
Similar cases happen when an unsupported intent is unexpectedly activated: in this case, Botium waits for a response that never arrives, eventually resulting in a timeout and the mutant being killed.

We also observed some cases of broken mutants produced by \tool, particularly when operating on conversational flows, since the resulting mutants can lead to invalid conversations that are unusable for chatbot training, or to contradictory rules/stories caused by missing interactions. In contrast, this behavior was not observed in our previous experiments with Dialogflow, where all the produced mutants were modified working copies of the original chatbots~\cite{ferdinando2024mutabot}. This difference might be due to a combination of the greater complexity of the mutated chatbots and the higher complexity of flow implementation. 



We can identify three sources of test weaknesses that result in high mutant survivability: (i) \textit{oracle imprecision}, (ii) \textit{oversimplified conversational test scenarios}, and (iii) \textit{limit in exploitable training data}.

Concerning the \textit{oracle imprecision} weakness, Botium is designed to detect intents but cannot precisely assert on bot responses; thus it misses all mutants that provide a wrong response in a correct intent. For instance, in a mutated version of the chatbot 
``Rock Papers Scissors'', removing the entity that stores the user's choice causes the bot responds with \textit{You chose None} when the user says \textit{Paper}, which is (erroneously) classified as a correct response by Botium. 

Concerning the \textit{oversimplified conversational test scenarios} weakness, in both the experiments with Dialogflow and Rasa, we observed how Botium generates very simple test scenarios, composed of only few user-bot interactions without expanding any followup cases. As a result, long or constrained flows (e.g., those activated by a precise user choice) are very rarely exercised, leading to a low mutation score (30\% in Dialogflow~\cite{ferdinando2024mutabot}, 23-58\% in Rasa from Table \ref{tab:results}). For example, Botium fails to detect a mutant of 
``PJs Chatbot'' in which the step asking for the delivery address after selecting the home delivery option has been removed, because the tool does not generate a test scenario exercising this constraint. Furthermore, Botium misses all mutants that affect session expiration time, since timing aspects are not explicitly addressed by the tool. 

Concerning the \textit{limit in exploitable training data} weakness, and in line with previous experiments~\cite{ferdinando2024mutabot,gomez2024mutation}, we observed how Botium is strongly affected by the training data, which serves as its sole source for test generation.
Thus, if a chatbot is not designed with training phrases that cover specific intents or entities (e.g., an entity value is not involved in any phrase), the generated test suite will completely miss them. This situation is particularly common for \textit{fallback intents} (i.e., the intents triggered when user input does not match any known intent with sufficient confidence), as they represent negative scenarios that are rarely covered in the training phrases~\cite{ferdinando2024mutabot}. 

The new findings targeting the Rasa platform strengthen our preliminary results obtained on Dialogflow, showing that state-of-the-art chatbot testing tools exhibit consistent weaknesses in revealing core defects across multiple platforms and conversational aspects, and highlighting the need for more advanced mechanisms to generate tests capable of exposing them. 

\section{Related Work} \label{sec:related}
With the growing popularity of chatbots in daily activities, several conversational testing approaches have recently been proposed~\cite{li2022review,lambiase2024motivations}. However, these are often constrained by the unique challenges of testing dialogue-based systems, such as the lack of suitable subjects, and the inherent difficulty of defining precise test oracles.

The most popular testing framework for task-based chatbots is Botium~\cite{botium}, which offers both open-source and commercial solutions, and produces tests as plain-text conversations in the form of user-bot interactions. Botium has served as the foundation for other testing proposals~\cite{bravo2020testing,canizares2024coverage,rapisarda2025test}, leveraging its test generation capability by augmenting test suites with both correct and perturbed data to test deeper aspects of software robustness and coverage. 
Guichard \textit{et al.}~\cite{guichard2019assessing} proposed a testing technique to build paraphrases from user requests. 
 Bozic \textit{et al.} addressed the oracle problem by introducing conversational input transformations and metamorphic rules, such as synonym substitution and word removal~\cite{bozic2019testing,bovzic2022ontology}. 

Although there is a growing number of chatbot testing proposals in the literature, there is still a lack of appropriate tools specifically aimed at analyzing faults in conversation-based software.
To address this gap, in a recent work we introduced \tool~\cite{ferdinando2024mutabot}, a tool that adapts mutation testing to the context of conversations and is designed to target faults occurring in multi-platform chatbots, originally implemented and evaluated for Dialogflow. In that study, we reported that Botium was able to detect only up to 37\% of the injected bugs. 

In parallel to our work, Gómez-Abajo \textit{et al.} proposed a set of 19 mutations for Dialogflow and Rasa chatbots~\cite{gomez2024mutation}, along with a mutation environment for their application~\cite{gomez2025wodel}. They reported similar findings, with Botium killing on average 46\% of mutants in Rasa.
Unlike our work, the proposed mutations mainly focus on element deletions, while our operators are designed to target possible finer-grained conversational aspects that can be more difficult for testing techniques to detect, 
as we observed in our preliminary experiment (e.g., replacing the name of an entity with another existing entity name, or extend/reduce the lifespan of a context variable, to simulate a programming mistake). 

\section{Conclusion} \label{sec:conclusion}
Mutation testing is an important approach to assess the effectiveness of test suites and test generation tools. Task-based chatbots represent a relevant domain that requires specific mutations to deal with the many bugs that might affect conversations. Since chatbots share similar conversational attributes across multiple domains and platforms, in this paper we present an extension of \tool, a conversational mutation technique that is platform-agnostic.


Chatbot mutation testing can help improve the state-of-the-art test generation techniques that may otherwise fail to reveal bugs in conversations, in particular when these bugs emerge only in long conversational and time‑dependent scenarios. In this paper, we discuss how we extended \tool to support Rasa chatbots in addition to Dialogflow chatbots. Early results show that test generation tools consistently exhibit similar weaknesses when generating test cases for both Dialogflow and Rasa chatbots. These weaknesses concern oracles, the complexity of dialogues, and the dependency on training data.

We are continuously extending the capabilities of \tool by covering additional conversational aspects and carrying out an extensive experimental evaluation across multiple platforms, while further investigating mutation testing of task‑based chatbots to gain deeper insights into the strengths and weaknesses of existing test generation tools.
We also plan to enhance the functionalities of \tool to automate error reporting during testing and equivalent mutant detection, assessing its scalability with highly complex chatbots.

{\small \emph{Acknowledgments:} This work has been partially funded by the Engineered MachinE
Learning-intensive IoT systems (EMELIOT) national research project,
which has been funded by the MUR under the PRIN 2020 program
(Contract nr. 2020W3A5FY).}


\bibliographystyle{IEEEtran}
\bibliography{IEEEabrv,references}

\end{document}